\documentstyle[12pt,moriond,epsfig]{article}
\begin{document}

\title{Scales and Phases in Particle Physics and Cosmology}
\author{R. Peccei \\ Department of Physics and Astronomy \\
University of California, Los Angeles, CA 90095-1547} 

\maketitle

\begin{abstract}

I describe and analyze the various energy scales that emerge from studying the structure of the Standard Models of particle physics and cosmology. Remarkably, save for the scale of the cosmological vacuum energy, all the scales below the Fermi scale appear to be naturally associated with generalized see-saw mechanisms. I also briefly discuss the known and presumed CP-violating phases emerging from physics associated with these models and their physical extensions. I end by commenting on the insights one might expect to gain from experiment on these issues in the next decade.

\end{abstract}

\section{Triumphs and Mysteries of the Standard Model}

In the last half century, particle physics and cosmology have gone 
from infancy to maturity with the establishment of their respective
Standard Models: the $SU(3)\times SU(2)\times U(1)$ gauge theories
of the strong and electroweak interactions;\cite{SM} and the hot
Big Bang model,\cite{BB} augmented by inflation.\cite{Guth}  These Standard
Models correlate and codify an enormous amount of data.  For
instance, direct measurements of the mass of the W-boson and of
the top quark correlate extremely well with the indirect determination of these quantities through electroweak radiative corrections,\cite{LEPEW} thereby helping to bound the mass of the Higgs boson in the electroweak theory to below 220 GeV at 95\% C.L.\cite{LEPEW}  The strong coupling constant of QCD, $\alpha_s(q^2)$, runs as predicted by the theory decreasing by a
factor of more than 4 as $q^2$ ranges from 1 GeV$^2$ to (200 GeV)$^2$.  On the cosmological side, the primordial abundances of ${}^4$He, deuterium, ${}^3$He and ${}^7$Li pin down the baryon to photon ratio in the Universe to a narrow range ($\eta = (3.8\pm 0.2)\times 10^{-10}$~\cite{BBN}) indicating that baryons contribute only at the few percent level to the closure density of the Universe.  The clear appearance of the first Doppler peak at $\ell\sim 200$ in the recent
data on the cosmic microwave background angular spectrum, obtained by Boomerang\cite{Boo} and Maxima,\cite{max} argue convincingly for a flat Universe ($\Omega = 1$), as predicted by inflation.\cite{inflation}

Although these and other triumphs of the Standard Models are impressive, perhaps what is more important about these models is that their establishment permits one to ask even deeper questions about the structure of the Universe and of physical law.  In so doing, one uncovers a set of mysteries whose resolution impels one to search for physics beyond that already codified in the Standard Models.  For particle physics, the list of mysteries is centered on the {\it origins of the Fermi scale} ($v \simeq 250$ GeV) characterizing
the electroweak breakdown
\begin{equation}
SU(2)\times U(1)\stackrel{v}{\rightarrow} U(1)_{\rm em}~,
\end{equation}
and on the {\it flavor problem}, connected to the origin of the fermion masses and of their mixing.  For cosmology, the list of mysteries is focused on {\it the composition of the Universe}, in particular on the detailed fractions which make up the Universe's energy density:
\begin{equation}
\Omega = \frac{\rho}{\rho_c} = \Omega_B + \Omega_{DM} + \Omega_\Lambda \stackrel{?}{=} 1~,
\end{equation}
and how the interplay of these various components leads to {\it the formation and evolution of structure in the Universe}.  

At a deep level, these mysteries cannot be unconnected, since the structure of the Universe emerges as a result of physical law.  However, these deep connections are still largely unknown, although one can envisage circumstances where experiment may make these connections apparent.  For example, if it turns out that (part of) the dark matter of the Universe are neutralinos whose mass results from SUSY breaking, one may be able to directly connect $\Omega_{DM}$ with $v$ since SUSY breaking, at least in certain models, can be the source of the breakdown of the electroweak theory.\cite{IR}

In this talk I would like to focus on a subset of the mysteries associated with the Standard Models: that of the energy scales and of the CP-violating phases which emerge from these models.  I concentrate on these issues because, aside from their intrinsic interest, the scales and phases present in the Standard Models very naturally force one to think of the interrelations which must exist between particle physics and cosmology.  I'll begin my discussion by talking about scales, as this helps to frame the problem, and return towards the end to the issue of phases.

The particle physics Standard Model has one clear {\it natural} scale, $\Lambda_{\rm QCD} \sim 300$ MeV, which is the scale associated with the running of $\alpha_s(q^2)$.  Roughly speaking, one can define $\Lambda_{\rm QCD}$ as the place where $\alpha_s(\Lambda^2_{\rm QCD}) = 1$.  The reason that this scale is natural is that $\Lambda_{\rm QCD}$ depends very weakly on a possible physical cutoff for the model set by gravity, whose natural scale is set by the Planck mass $M_P \sim 10^{19}$ GeV fixed by the strength of the Newtonian constant, $G_N = 1/M_P^2$.  The strong coupling constant $\alpha_s(q^2)$ only changes logarithmically with the scale $q^2$, so effectively $\Lambda_{\rm QCD}$ is fixed by knowing the value of $\alpha_s(M_P^2)$ and not directly by the Planck mass itself.  That is, $\Lambda_{\rm QCD}$ is not directly proportional to $M_P$, but only logarithmically dependent on this scale.

It is possible that the Fermi scale $v$ is also a natural scale in the same sense as $\Lambda_{\rm QCD}$.  For this to be the case, one has to imagine that $v$ is related to the running of the coupling constant of some other dynamical theory responsible for the electroweak breakdown.  This is what emerges in Technicolor theories\cite{Technicolor} where the underlying Technicolor coupling constant becomes strong at a scale of order $v$, $\alpha_{\rm TC}(v^2) = 1$, and $v$ characterizes the scale of the fermion condensates in the theory, $\langle\bar TT\rangle\sim v^3$, which lead to electroweak breakdown.  In my view, however, this is an unlikely scenario. As I mentioned earlier, electroweak radiative corrections are consistent with a relatively light Higgs boson, while  typically, Technicolor theories
are associated with a TeV scale Higgs boson. Thus to reproduce the radiative correction results one must somehow arrange to cancel these contributions with those coming from other parts of the theory, so as to mimic the effects of a light Higgs boson.\cite{ABC}  Although this may be possible, what proves even harder in these theories is to be able to consistently generate  the full spectrum of quarks and leptons dynamically with new physics at or near the Fermi scale.\cite{PecceiLL}

What is more likely is that the Fermi scale $v$ is associated with the vacuum expectation value (VEV) of some elementary scalar field.  To render this scale natural, however, there must exist some protective supersymmetry, for otherwise radiative corrections would cause a shift in $v$ proportional to the scale of the cutoff.\cite{natural} Supersymmetry, even if broken, renders the dependence on the cutoff only logarithmic.  This is illustrated schematically below for the shift in $v^2$ in theories where there is not, or there is, a supersymmetry. One finds:
\begin{equation}
\Delta v^2 \sim \left\{ \begin{array}{ll}
\alpha M_P^2 & {\rm unprotected} \\
\alpha(m^2-\tilde m^2) \ln M_P/v & {\rm protected~ by~ SUSY}
\end{array}
\right.
\end{equation}
In the above, $\tilde m$ denotes generically the mass of the SUSY partners of a particle of mass $m$.

I should note that if $v$ is associated with the VEV of some scalar field, then the Yukawa couplings of this scalar field (or fields) to fermions provide a mechanism for generating fermion masses and mixing.  Although fermion masses are proportional to $v$, their actual values depend on these Yukawa couplings
\begin{equation}
m_f \sim h_f~v~.
\end{equation}
Because the couplings $h_f$ are unknown, Eq. (4) does not explain the origin of fermion masses.  However, the physics which determines these Yukawa couplings can very well originate at a scale much larger than the Fermi scale.  So, in contrast to Technicolor theories, one effectively postpones the resolution of the origins of the mass and mixing structure of fermions to a deeper level of physics.  Ignorance here, although not quite bliss, allows one to separate the problem of the origin of the Fermi scale from that of the origin of fermion masses---  a real advantage.

\section{A Plethora of Scales: Facts and Fancy}

Besides $v\sim 250$ GeV and $\Lambda_{\rm QCD} \sim 300$ MeV and the scales associated with the fermion masses $(m_e \sim 0.5~{\rm MeV},\ldots, m_t\sim 170~{\rm GeV})$, two new scales have emerged from experiment recently:
\begin{description}
\item{i)} The SuperKamiokande Collaboration\cite{SuperK} has found evidence for oscillations of neutrinos produced in the atmosphere, typified by a mass difference squared around $\Delta m^2\sim 2.5\times 10^{-3}~{\rm eV}^2$.  Assuming that one of the neutrinos dominates $\Delta m^2$, this indicates a mass for this neutrino of order
\begin{equation}
m_\nu \sim 5\times 10^{-2}~{\rm eV}~.
\end{equation}
Furthermore, hints of solar and accelerator neutrino oscillations\cite{oscl} suggest the existence of other neutrino masses also in the sub-eV or eV range.
\item{ii)} Studies of supernovas at large redshift\cite{SN} provide evidence for an {\it accelerating} Universe.  The measured deceleration parameter
\begin{equation}
2q_o = \Omega_M - 2\Omega_\Lambda
\end{equation}
turns out to be negative, rather than positive, hinting at the existence of a {\it dark energy} component in the Universe.  For $\Omega_\Lambda \sim 0.7$, as indicated by the data, the corresponding vacuum energy density also appears to involve a sub-eV energy scale:
\begin{equation}
\rho_{\rm vacuum} \sim (3\times 10^{-3}~{\rm eV})^4~.
\end{equation}
\end{description}
Clearly the energy scales associated with neutrino masses, Eq. (5), and the vacuum/dark energy in the Universe, Eq. (7), signal phenomena that are {\it beyond} the Standard Model, precisely because of the discrepancy of these scales with those in the Standard Model.

Experiment, however, is not the only source of new scales.  There are a number of theoretical arguments which introduce new scales, beyond those present in the Standard Model and the scale of gravity, $M_P$.  Perhaps the most famous of these ``theoretical" scales is the GUT scale, $M_G$, signalling the unification of the Standard Model forces into one.  A variety of GUT groups $G$ have been suggested ($SU(5)$,
$SO(10)$, $E_6$, etc.)\cite{GUT} with $M_G$ being the scale at which the group $G$ breaks down spontaneously to the Standard Model group
\begin{equation}
G\stackrel{M_G}{\longrightarrow} SU(3)\times SU(2)\times U(1)~.
\end{equation}
Evidence for $M_G\sim 2\times 10^{16}$ GeV comes from the evolution of the Standard Model coupling constants, assuming the existence of
SUSY matter above a TeV.\cite{Amaldi}

The GUT scale $M_G$ is only one example of a large scale inferred
from theory.  There are literally hosts of other large scales, associated with different new physics phenomena, which have been proposed.  Some selected examples include:
\begin{description}
\item{i)} The scale $f_a$ of the spontaneous breaking of the chiral symmetry $U_{\rm PQ}(1)$ introduced to ``solve" the strong CP problem.\cite{PQ}  This symmetry has an associated very light particle with it, the axion,\cite{WW} whose mass is inversely proportional to $f_a$.  Bounds from astrophysics and cosmology restrict the axion mass to a rather narrow range $[10^{-6}~{\rm eV} \leq m_a \leq 10^{-3}~{\rm eV}]$, which give a range for $f_a$:\cite{bounds}
\begin{equation}
10^9~{\rm GeV} \leq f_a \leq 10^{12}~{\rm GeV}~.
\end{equation}
\item{ii)} The scale of the inverse compactification radius, $R^{-1}$, in theories with $d>4$ space-time dimensions.  In these theories\cite{Dimopoulos} one identifies the Fermi scale $v$ with the scale $M$ of gravity in $(d+4)$ dimensions, with $R$ being the size of the compact dimensions.  By comparing Newton's law in $(d+4)$
and in 4 dimensions, one identifies
\begin{equation}
M_P = M(MR)^{d/2}~,
\end{equation}
so that for $d=2$, $R=1$ mm (and $R^{-1} = 2\times 10^{-4}~{\rm eV}$),
while for $d=6$, $R = 10^{-13}$ cm (and $R^{-1} = 200$ MeV).  Obviously these theories give modifications to the usual gravitational potential for distances $r\sim R$.
\item{ii)} The scale $\Lambda_s$ of spontaneous breakdown of supersymmetry in supersymmetric extensions of the Standard Model.
If SUSY is broken, as it is often assumed,  in a hidden sector coupled to matter only by gravitational interactions,\cite{SUGRA} then the masses $\tilde m$ of the SUSY partners of ordinary particles are given by $\tilde m\sim \Lambda_s^2/M_P$.  Assuming that $\tilde m\sim v\sim~{\rm TeV}$, this fixes the scale of SUSY breaking:
\begin{equation}
\Lambda_s\sim 10^{11}~{\rm GeV}~.
\end{equation}
\end{description}

It is not clear whether any of the above ``theoretical" scales really obtain in nature.  Nevertheless, it is useful to display all the scales we discussed in a figure (Fig. 1).  In what follows, I will use this figure in conjunction with the idea of natural {\it see-saw mechanisms}, to try to understand better the nature of the different scales encountered.

\begin{figure}[t]
\center
\epsfig{file=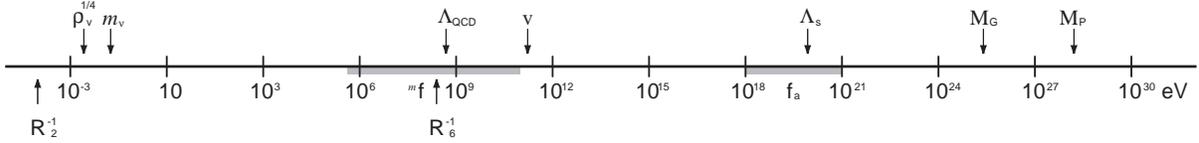,height=0.75in}
\caption{Scales entering in particle physics and cosmology}
\end{figure}

\section{Light States and Their See-Saw Windows}

To proceed, it is important to make a distinction between {\it physical scales} and {\it masses of physical states}.  In a dynamical theory like QCD these are closely interrelated.  Because at scales of order $\Lambda_{\rm QCD}$ the strong coupling constant becomes strong, it is perfectly natural that the mass of the hadrons should also be of order $\Lambda_{\rm QCD}$:
\begin{equation}
M_{\rm hadrons} \sim \Lambda_{\rm QCD}~.
\end{equation}
This is no longer the case for the masses of quarks and leptons.  Although these masses are all proportional to the Fermi scale, the wide range of Yukawa couplings in Eq. (4) $[2\times 10^{-6}\leq h_f \leq 1]$ give a strong hint that the origin of the flavor masses most probably originates at a scale much above $v$.

I believe it is even clearer that the origin of the extraordinary small masses  of neutrinos, or of other postulated very light states, is connected with physics at a very much higher scale than the Fermi scale. This inference emerges as the result of a dynamical mechanism, known colloquially as the {\it see-saw mechanism}.\cite{YGRS}  Although the
argument connected with the see-saw mechanism is by now rather familiar, let me touch upon it briefly.  Because neutrinos are neutral, it is possible for these states to have three different types of mass terms, only one of which conserves lepton number.  For the charged quarks and leptons only a (Dirac) mass term is allowed by charge conservation.  However for neutrinos, if  one does not impose lepton number conservation, \footnote{Because there is no sacrosant reason for lepton number to be conserved, one should be open to the possible existence of these Majorana mass terms.} one can countenance the presence of (Majorana) mass term involving two neutrinos, or two anti-neutrino fields.  

As a result, the most general mass terms for neutrinos, considering for simplicity only one neutrino flavor, involves three different mass parameters:\cite{PecceiM}
\begin{equation}
{\cal{L}}_{\rm mass} = m_D\bar\nu_L\nu_R + \frac{1}{2} m_R\nu_R^TC\nu_R + \frac{1}{2} m_L\nu_L^TC\nu_L + {\rm h.c.}
\end{equation}
The Dirac mass term $m_D$ is proportional to the Fermi scale $v$, since it violatess the electroweak $SU(2)$ symmetry by 1/2 a unit.  Since right-handed neutrinos are $SU(2)$ singlets, the Majorana mass $m_R$ is {\it independent} of the electroweak symmetry breaking scale $v$.  The other Majorana mass parameter $m_L$, on the other hand, violates the electroweak $SU(2)$ by a full unit and hence $m_L\sim v^2$. 

If one assumes there are no right-handed neutrinos in the theory, then the neutrino mass $m_\nu$ can only be due to $m_L$.  For neutrinos to have a very small mass it must be then that $m_L$ is inversely proportional to a large mass scale $M\gg v$,
\begin{equation}
m_\nu \equiv m_L \sim \frac{v^2}{M}~.
\end{equation}
Alternatively if right-handed neutrinos exist, one can also obtain
light neutrino masses if the Majorana mass $m_R$ is very large.\cite{YGRS}  Neglecting for simplicity $m_L$, the neutrino mass matrix in this case is given by the $2\times 2$ structure
\begin{equation}
{\cal{M}} = \left(
\begin{array}{cc}
0 & m_D \\ m_D & m_R 
\end{array}
\right)~.
\end{equation}
If $m_R \gg m_D$ the eigenvalues of this matrix imply the existence of both a superheavy neutrino of mass $m_R$ and a superlight neutrino of mass
\begin{equation}
m_\nu \simeq \frac{m_D^2}{m_R} \sim \frac{v^2}{m_R}~.
\end{equation}

It is clear that, in either case, one gets the same {\it see-saw}
formula in which superlight neutrinos arise as the result of the existence of a large scale.  To obtain neutrino masses of the order of what one infers from the SuperKamiokande data ($m_\nu \sim 5\times 10^{-2}$ eV) requires
\begin{equation}
m_R~~ \mbox{and/or}~~ M \sim (10^{11}-10^{15})~{\rm GeV}~.
\end{equation}

Neutrino masses are not the only examples of a see-saw mechanism at work.  Similar formulas emerge for axions, with the mass of the axion being driven below $\Lambda_{\rm QCD}$---the scale associated with the anomaly in the $U(1)_{\rm PQ}$ current---by the ratio of $\Lambda_{\rm QCD}$ to the scale $f_a$ of $U(1)_{\rm PQ}$ spontaneous
breaking, $m_a \sim \Lambda_{\rm QCD}^2/f_a$.  We saw earlier that also the masses of sparticles have these see-saw forms: 
$\tilde m\sim \Lambda_s^2/M_P$, with $\tilde{m}$ being driven below $\Lambda_s$ by the ratio of the scale of supersymmetry breaking to the Planck mass, since in these theories \cite{SUGRA} the symmetry breaking sector only  communicates with the visible sector through gravitational interactions.  One of the nice features of this scenario of supersymmetry breaking is that, as a result of the large top Yukawa coupling, one can induce electroweak breaking from SUSY breaking.\cite{IR}  So the Fermi scale itself is also given, effectively, by a see-saw formula
\begin{equation}
v\sim \frac{\Lambda_s^2}{M_P}~.
\end{equation}

All of these see-saw formulas raise the question of whether all observable physics has its origin at high scales, near the Planck mass.  Although one can make arguments for this to be the case for almost all the relatively light scales (scales at or below the Fermi scale), the scale associated with the vacuum energy seems to belie this assertion.  This scale, as we discussed [c.f. Eq. (7)], is tiny.  Furthermore, this cosmological vacuum energy is not really identifiable with the vacuum energy one encounters in particle physics, which is related to the vacuum expectation value of the trace of the energy momentum tensor $\langle T_\mu^\mu\rangle$. The different pieces of the Standard Model contribute to  $\langle T_\mu^\mu\rangle$  with the typical scale one associates with each of these components.  Thus
\begin{equation}
\langle T_\mu^\mu\rangle_{\rm QCD} \sim\Lambda^4_{\rm QCD}~; ~~~~
\langle T_\mu^\mu\rangle_{\rm EW} \sim v^4
\end{equation}
which lead to values for the vacuum energy enormously larger than the value given in Eq. (7).

To avoid this conundrum, which is essentially the mystery of why the cosmological constant either vanishes or nearly vanishes,\cite{cosmo} many physicists recently \cite{Q} have tried to interpret $\rho_{\rm vacuum}$ not as a vacuum energy but as some form of {\it dynamical dark energy} associated with the potential energy of a scalar field Q.  This, so called, quintessence field has negative pressure and becomes dynamical close to present times.  Hence
\begin{equation}
\rho_{\rm vacuum} \sim V(Q)~.
\end{equation}
For this to happen, however, the mass of quintessence must be extraordinarily tiny, with $m_Q$ being of the order of the Hubble constant now.  Hence
\begin{equation}
m_Q \sim H_o \sim\left[\frac{V(Q)}{M_P^2}\right]^{1/2} \sim
\frac{\rho_{\rm vacuum}^{1/2}}{M_P}\sim 10^{-31}~{\rm eV}~.
\end{equation}
Although another see-saw formula emerges in Eq. (21), this see-saw does not 
connect $\rho^{1/4}_{\rm vacuum}$ to some large scale.  Rather Eq. (21) introduces a superlight scale for quintessence much below
$\rho^{1/4}_{\rm vacuum}$!  This is very troubling, since it is
difficult to understand physically how one can naturally  keep $m_Q$ so light, without essentially decoupling quintessence from the visible sector.\cite{KL}

\section{Phases: Known, Suspected, and Unwanted}

I would like now to make some comments on the phases which enter in the Standard Models of particle physics and  cosmology.  In the usual three-generation minimal Standard Model of particle physics, the CKM model,\cite{CKM} there is only {\it one} CP-violating phase $\gamma$ associated with the quark mixing matrix.  In addition, there is an {\it effective vacuum angle}
$\bar\theta$ which also leads to CP-violating effects.  However, to avoid contradiction with the strong bounds on the electric dipole moment for the neutron, this CP-violating parameter must be very small, $\bar\theta < 10^{-10}$.\cite{RDP}  One must adduce reasons, like the existence of a $U(1)_{\rm PQ}$ symmetry, to explain this latter fact.  In contrast, one knows from experiment that the CP phase $\gamma$, if the standard CKM model\cite{CKM} is correct, is
large.  CKM analyses done in preparation for the turn-on of the SLAC B-factory \cite{BABAR} are consistent with
\begin{equation}
45^\circ < \gamma < 135^\circ~,
\end{equation}
with more recent analyses being even more restrictive [c.f.\cite{Stocchi} which gives $\gamma = (56\pm 7)^\circ$].

It is important to note that, even though the phase $\gamma$ in the CKM model appears to be large, the model nicely explains why one should expect the CP-violating effects in the Kaon system to be small.  As is well known, this arises because the phase $\gamma$ enters physically only if all 3 generations of quarks are involved in the process in question.  In the Kaon system this happens only through virtual loops and the resulting amplitudes are small due to the suppression of interfamily mixing in the CKM matrix.
This suppression, however, can be avoided in B-decays. Hence,  one of the most robust predictions of the CKM model is that in certain B-decay processes---like the decay of $B_d$, or $\bar B_d$ to $\psi K_s$--- CP-violating effects should be large. In particular, the CKM model predicts that the CP-asymmetry measured in the time development of a state originating as a $B_d$ and decaying into $\psi K_s$ should be quite significant, with the relevant CP-violating parameter
$\sin2\beta$ estimated to be around 0.7-0.8.\cite{Stocchi}  Interestingly, although early results from CDF seemed to be centered around this value ($\sin2\beta = 0.79 \pm 0.42$\cite{CDF}), albeit with large errors, new data emerging from the B-factories at SLAC and KEK reported at the Osaka Conference have a much smaller central value ($\sin2\beta =0.12 \pm 0.37 \pm 0.09$\cite{SLAC}; $\sin2\beta =0.45 \pm 0.44 \pm 0.08$\cite{KEK}),
again however with big errors.  Clearly more data is needed before one can make a judgement on the validity of the CKM model for describing
CP-violation in the quark sector.

Irrespective of how the above question is resolved, it is clear that if neutrino have mass one should expect additional CP-violating phases associated with the lepton mixing matrix.  These phases, however, are likely to be even harder to dig out.  Just as the effects of the CKM phase $\gamma$ are not observable if all three generations of quarks are not involved, no CP-violating effects due to the lepton mixing matrix are observable if the neutrinos are degenerate in mass.  It follows, therefore, that CP-violating effects in the neutrino sector are proportional to neutrino mass differences, and thus are intrinsically small since the neutrino masses themselves are so tiny.

These CP-violating phases in the neutrino sector, however, are not the only other CP-violating phases besides $\gamma$ which are likely to exist.  There is a general theorem\cite{RDPWIN} that states that if there is CP violation, then all parameters in the Lagrangian that can be complex must necessarily be so to preserve the renormalizability of the theory.  Theories which involve more fields than those present in the Standard Model, as is the case for supersymmetry, necessarily have a larger number of parameters in the Lagrangian,  and so end up by having a number of new CP-violating phases.  Thus, it is symptomatic of theories beyond the Standard Model of particle physics that these theories necessarily involve additional CP-violating  phases beyond the single CKM phase $\gamma$.

The same can be said for cosmology.  As soon as one tries to incorporate more phenomena into the Big Bang plus inflation Standard Model of cosmology, one ends up by introducing new sources for CP violation.  A good example is provided by baryogenesis, whose understanding in cosmology requires the existence of new phases beyond the CKM phase $\gamma$.  That this is the case is clear for GUT baryogenesis, since GUT models always contain further phases beyond $\gamma$ associated with GUT interactions.\cite{RDPBlois}  However, this is also the case for electroweak baryogenesis, since the standard CKM scenario does not give a sufficiently large asymmetry because of the very strong interfamily suppression.\cite{HN}  In addition,  the electroweak transition in the Standard Model is not strongly enough first order\cite{Buch} to prevent most of the produced baryon asymmetry to be washed away.  Remedies to these problems, usually involving supersymmetry,\cite{Carena} always end up by invoking theories with more degrees of freedom and thus more CP-violating phases.  At any rate, these cosmological considerations already provide compelling evidence for physics beyond the particle physics Standard Model.

Many of the new phases which enter in extensions of the Standard Model of particle physics lead to {\it flavor conserving} CP-violating effects.  Such CP-violating phases are more difficult to pin down than the CKM phase $\gamma$.  The best places to look for these new phases are in places like the neutron electric dipole moment, or the transverse muon polarization in $K_{\mu 3}$ decays, where the contributions from the CKM phase $\gamma$ is either small or vanishes.  Unfortunately, we still do not have any evidence to date for the presence of any CP-violating, flavor conserving phases. Nevertheless, theoretically we fully expect them to be there!

A final comment on phases is well worth making here.  Often it turns out that extensions to the Standard Model of particle physics are associated with the existence of new CP-violating phenomena which, in certain cases, turn out to give huge effects.  Because such effects are not seen experimentally, model builders often have to work hard to adduce reasons why these extra CP-violating phases must be very small.  I spoke earlier of one such example, connected with the strong CP-problem, where the CP-violating parameter $\bar\theta$, associated with the vacuum angle of the theory had to be very small [$\bar\theta < 10^{-10}$] so as not to violate the existing very strong limits on the neutron edm.  A similar situation occurs in most simple supersymmetric extensions of the Standard Model.  These theories contain additional CP-phases which, as it turns out, contribute substantially to the neutron edm.  To avoid troubles with experiment, if the masses of the superpartners are below a TeV or so, then the pure SUSY CP-phases must be assumed to be rather small [$\tilde\gamma\leq 10^{-2}$\cite{Hall}].  Why should one have such small phases in these models is not clear.  These and other unwanted CP-violating phases, more than anything else, are good evidence that we still do not understand at a deep level the structure of, and the rationale for, the CP-violating phases which are present in nature.

\section{Experiment to the Rescue}

Many of the issues I have discussed in this talk are not easily clarified because forthcoming experiments which could shed some light on these issues have necessarily limited scope in sensitivity and energy range.  Nevertheless, much will be learned in the coming decade.

First of all, with the LHC, if not sooner, we will be able to understand the nature of the Fermi scale.  In particular, if low energy supersymmetry is discovered, then the connection with current ideas for physics at the Planck scale based on superstrings will be strongly boosted.  At the same time, experiments with neutrinos--- solar, atmospheric, long- and short-baseline--- will help map the mass and mixing matrix structure of the neutrino sector, thereby opening up a further window onto the physics at high mass scales, otherwise unreachable by other means.

Much will be learned also about CP-violation in B-decays.  It is quite possible that through these experiments we may discover other sources of CP-violation besides the CKM phase $\gamma$.  If so, this may help make the connection with baryogenesis more directly, although unravelling how baryogenesis itself proceeds is a great challenge.  
Amusingly, there is presently a strong resurgence of the idea put forth more than a decade ago by Fukugita and Yanagida\cite{FY} that neutrinos are the key to establishing a non-trivial baryon asymmetry in the Universe.  In their scenario, lepton violating processes, associated with super massive neutrinos, build up a lepton asymmetry in the early Universe which is transformed into a baryon asymmetry at or above the electroweak phase transition. So phases in the neutrino sector may be the key to our own existence!

Particle physics experiments, however, are not likely to be the only sources for new insights.  Indeed much excitement is expected from a host of new experimental and observational efforts in cosmology.  Of particular importance are the upcoming missions [MAP, Planck] which will map the temperature angular anisotropy of the cosmic microwave background with great precision, thereby helping to pin down a host of cosmological parameters including $\Omega_\Lambda$.  Of course, if we are lucky, it is possible that there will be direct detection of the dark matter which dominates the matter content of the Universe.  This could be of enormous help to clarify what scales and associated physics are connected with this phenomenon.  Is the physics scale associated with dark matter one of the scales sketched in Fig. 1, or is it an entirely new scale?

Another important piece of cosmological evidence may emerge by studying the structure of the Universe at various large red shifts.  If one is really successful in doing this, it may be possible to trace the evolution of $\Omega_\Lambda/\Omega_M$ and hence to ascertain whether $\rho_{\rm vacuum}$ is a static quantity (i.e. due to a cosmological constant) or is more dynamic (i.e. due to something like quintessence).

Clearly it would be fun to continue to speculate on the insights which future results in both particle physics and cosmology will bring.  However, it is probably more prudent to let future experiments speak for themselves !  
                                           
\section*{Acknowledgements}

I would like to thank J. Tran Thanh Van for the splendid
hospitality in Hanoi.  This work was supported in part by the Department of
Energy under Contract DE-FE03-91ER40662, Task C.

\end{document}